\documentclass[aps,prb,twocolumn,superscriptaddress,floatfix,longbibliography]{revtex4-2}

\usepackage{graphicx,amsmath,amssymb,epstopdf,wasysym}
\usepackage[space]{grffile}
\usepackage[dvipsnames]{xcolor}
\usepackage{braket,amsfonts}
\usepackage{dsfont}
\usepackage{cancel}
\usepackage{comment}

\usepackage{hyperref}
    \hypersetup
    {   
        unicode=true,           
        pdftoolbar=true,        
        pdfmenubar=true,        
        pdffitwindow=false,     
        pdfstartview={FitH},    
        pdfauthor={Ivan Pasqua, Gabriele Bellomia, Bartomeu Monserrat, Michele Fabrizio, Carlos Mejuto-Zaera},
        pdftitle={TopoGhost},
        pdfkeywords={topological insulators, topological transitions, magnetic ordering, quantum embedding, quasiparticle hamiltonian, ghost Gutzwiller},
        pdfnewwindow=true,      
        colorlinks=true,        
        linkcolor=MidnightBlue,        
        citecolor=Maroon,        
        filecolor=Brown,        
        urlcolor=Maroon          
    }

\newcommand{\up}{\uparrow}
\newcommand{\down}{\downarrow}
\newcommand{\be}{\begin{equation}}
\newcommand{\ee}{\end{equation}}
\newcommand{\bk}{\mathbf{k}}
\newcommand{\bR}{\mathbf{R}}
\newenvironment{eqs}%
{\begin{equation} \begin{aligned}}%
{\end{aligned} \end{equation} }
\newcommand{\beal}{\begin{eqs}}
\newcommand{\eal}{\end{eqs}}
\newcommand{\eqn}[1]{(\ref{#1})}
\newcommand{\dagga}{{\phantom{\dagger}}}

\newcommand{\fract}[2]{\frac{\displaystyle #1}{\displaystyle #2}}
\newcommand{\bw}{\begin{widetext}}
\newcommand{\ew}{\end{widetext}}
\newcommand{\ep}{\epsilon}

\newcommand{\bealn}{\beal\nonumber}

\begin{document}

\title{Band structure picture for topology in strongly correlated systems \\ with the ghost Gutzwiller ansatz}

\author{Ivan Pasqua}
\email{ipasqua@sissa.it}
\affiliation{International School for Advanced Studies (SISSA), Via Bonomea 265, I-34136 Trieste, Italy} 

\author{Antonio Maria Tagliente}
\affiliation{International School for Advanced Studies (SISSA), Via Bonomea 265, I-34136 Trieste, Italy}

\author{Gabriele Bellomia}
\affiliation{International School for Advanced Studies (SISSA), Via Bonomea 265, I-34136 Trieste, Italy} 

\author{Bartomeu Monserrat}
\affiliation{TCM Group, Cavendish Laboratory, University of Cambridge, J.J. Thomson Avenue, Cambridge CB3 0HE, United Kingdom} 
\affiliation{Department of Materials Science and Metallurgy, University of Cambridge, 27 Charles Babbage Road, Cambridge CB3 0FS, United Kingdom} 

\author{Michele Fabrizio}
\affiliation{International School for Advanced Studies (SISSA), Via Bonomea 265, I-34136 Trieste, Italy} 

\author{Carlos Mejuto-Zaera}
\email{cmejutozaera@irsamc.ups-tlse.fr}
\affiliation{Univ Toulouse, CNRS, Laboratoire de Physique Théorique, Toulouse, France.}

\date{\today}

\begin{abstract}
Understanding the interplay between electronic correlations and band topology remains a central challenge in condensed matter physics, primarily hindered by a language mismatch problem. While band topology is naturally formulated within a single-particle band theory, strong correlations typically elude such an effective one-body description. In this work, we bridge this gap leveraging the ghost Gutzwiller (gGut) variational embedding framework, which introduces auxiliary quasiparticle degrees of freedom to recover an effective band structure description of strongly correlated systems. This approach enables an interpretable and computationally efficient treatment of correlated topological phases, resulting in energy- and momentum-resolved topological features that are directly comparable with experimental spectra. We exemplify the advantages of this framework through a detailed study of the interacting Bernevig-Hughes-Zhang model. Not only does the gGut description reproduce established results, but it also reveals previously inaccessible aspects: most notably, the emergence of topologically nontrivial Hubbard bands hosting their own edge states, as well as possible ways to manipulate these through a finite magnetization. These results position the gGut framework as a promising tool for the predictive modeling of correlated topological materials.

\end{abstract}

\maketitle

\section{Introduction}
Among the key guiding principles in modern condensed matter physics, band structure topology (BST) and strong electronic correlations (SEC) have emerged as central paradigms for understanding and designing materials with tailored conducting and magnetic properties. Topology enables the design of robust quantum phases, resilient to disorder and external perturbations \cite{Hasan2010, RMPZhang}, while SEC governs the emergence of theoretically and technologically relevant phenomena such as high-temperature superconductivity \cite{Lee2006}, metal-Mott insulator transition \cite{RMPDMFT, Imada1998}, and colossal magnetoresistance \cite{Tokura2006, Vaz2010, Thle2020}.
Furthermore, strongly correlated phases of matter are arguably characterized by their tunability, i.e., the fact that slight modifications of external parameters like pressure, physical or chemical, doping or temperature result in dramatic changes of the electronic structure \cite{Dagotto2005, Kotliar2004}. 

In the last decade, the interplay between strong correlations and band-structure topology has attracted intense research efforts, entailing a wide variety of analytical and numerical methods \cite{Lu2013,Nandy2024,Hohenadler_REVIEW,Rachel_Review,Crippa_WeylDMFT,Crippa_WeylDMFT,Held_Weyl_LDA+DMFT,Gilardoni_manybodyZ2,Xu_2024_TenNetTopo, Bisht2024}. Yet, a unified theoretical framework able to address both aspects remains elusive, despite being crucial for the design of complex quantum materials. Such a framework would not only enable the stabilization of exotic and robust quantum phases, but also support their controlled realization in device-oriented settings. The urgency of this goal, which remains arguably an open challenge in the field, is underscored by a growing number of experimental breakthroughs, including the realization of chiral Kondo lattices \cite{GuerciScience} and signatures of unconventional superconductivity near reentrant and fractional quantum anomalous Hall insulator in moiré transition metal dichalcogenides \cite{arXivLi}, as well as evidence of chiral superconductivity in rhombohedral graphene \cite{rhombohedral_graphene_superc}, just to name a few.

Topological phases in solid-state systems have been widely explored for periodic models of noninteracting electrons, where physical observables are related to topological invariants that can be directly computed from the eigenstates of single-particle Hamiltonians \cite{Haldane1988, KaneMele, BHZmodel}, which are easy to diagonalize. However, this framework breaks down in the presence of electronic correlations, when the size of the Hamiltonian matrix scales with the Hilbert space rather than the system volume. In weakly correlated systems, effective single-particle descriptions, such as Density Functional Theory (DFT) \cite{Kohn1996,Jones2015,Martin2020}, can offer invaluable insights for guiding material design \cite{Slager2013,Kruthoff2017,Bradlyn2017,Antimo_topoDFT,Vergniory2019}, but they can fall short when interaction effects dominate over the band energy gain.  Conversely, state-of-the-art methods for strongly correlated systems, such as Dynamical Mean-Field Theory (DMFT) \cite{RMPDMFT,Kotliar2006,Paul2019}, excel at describing local correlation effects, but lack a natural and directly accessible band-structure picture applicable at both low and high energy scales, which is important to describe the inter-band coherence effects relevant to topology.
Moreover, DMFT and similar many-body methods can be computationally demanding across wide parameter regimes, critically hindering the extensive exploration of potential material candidates. This limitation is particularly relevant in the context of twisted material platforms, where the high degree of tunability underscores the richness of the phase diagram \cite{Li2021, Devakul2021, Kim2023}.

The recently introduced ghost Gutzwiller (gGut) framework~\cite{Lanata2017,guerci2019,frank2021,Mejuto2023a,Lee2023a,Lee2023b,Guerci2023} offers a promising new avenue to overcoming this challenge. This approach models correlation by applying local projectors to the ground state of a quasiparticle (QP), noninteracting Hamiltonian. By introducing auxiliary “ghost” degrees of freedom, the gGut method captures correlation-induced spectral features that are typically missed by independent-particle approximations, while achieving an accuracy comparable to single-site DMFT at a significantly lower computational cost \cite{lanata2015, Mejuto2024}. The resulting QP Hamiltonian captures SEC effects beyond the mean-field level, while retaining a band-theoretical description. In this context, we propose that the corresponding quasiparticle band structure enables the direct application of symmetry-based indicators \cite{Fu2007, Song2018} and all the other tools originally developed to characterize the topology of noninteracting systems \cite{Haldane1988, BHZmodel}. This, in turn, opens the possibility to straightforwardly calculate topological invariants and directly analyze edge states in open-boundary geometries, making gGut a powerful and intuitive tool for investigating topological phases in strongly correlated systems. 

In this paper, we introduce the use of gGut for treating correlated topological models, and exemplify its figures of merit with a comprehensive analysis of the interacting BHZ model.
This model case shows how the quasiparticle band picture reveals heretofore unobserved features of the electronic structure topology: besides recovering the established phenomenology of low-energy topology in the model, we observe that the Hubbard bands can become topologically nontrivial, presenting characteristic edge states.
Moreover, we show how the topological nature of the Hubbard bands can be controlled by leveraging magnetism, clearly exemplifying how the gGut description can excel at describing the tuneability of strongly correlated matter.
While our results here focus primarily on lattice systems, we hint at the promise for simulations in finite-size configurations, showing some sample calculations in layered structures, with clear implications for applications in technologically relevant scenarios.

\section{Model and Methods}
In the following, we briefly introduce our minimal yet paradigmatic model for interacting topological systems: the Bernevig-Hughes-Zhang (BHZ) model in presence of a multi-orbital repulsive interaction between electrons on the same site. This model captures the key ingredients underlying interaction effects on quantum spin Hall insulating states~\cite{BHZ_AFMTI_DMFT/NRG,BHZ_AFMTI_VCA,Sangiovanni2013, BHZAdriano1, BHZAndrea, BHZAdriano2, BHZCrippa, MottZerosGiorgio, BHZ_AFMTI_DMRG, BHZFrancesca, BHZIvan}. 

Then, we move forward to introduce the essentials of the quasiparticle picture lying at the core of the
gGut method, we relate it to the concept of Fermi-liquid quasiparticle 
Green's function~\cite{Fabrizio2022,Fabrizio2023,AndreaTopoZeros} and we show how the topology of the interacting model can be fully encoded in the gGut quasiparticle Hamiltonian, living in the auxiliary extended Hilbert space.

\subsection{Interacting Bernevig-Hughes-Zhang model}
\label{sec:Interacting BHZ model}
In this work, we focus on the BHZ model for the quantum spin Hall insulator (QSHI) phase of HgTe quantum wells \cite{BHZmodel}. This model involves two spinful Wannier orbitals per unit cell, one that transforms like $s$-orbitals, $\ket{\ell=0,\sigma}\equiv \ket{s \,\sigma}$, $\sigma=\up,\down$ being the spin projection along $z$, and the other like the $J=3/2$, $J_z=\pm 3/2$ spin-orbit 
coupled combinations of $p$-orbitals, 
$\ket{\ell=1,\ell_z=+1,\up}\equiv \ket{p \up}$ and 
$\ket{\ell=1,\ell_z=-1,\down}\equiv \ket{p \down}$.
We introduce two sets of Pauli matrices, $\sigma_a$ and $\tau_a$, $a=0,\dots,3$, with $a=0$ denoting the identity, which act, respectively, in the spin ($\up$, $\down$), and orbital ($s$, $p$), spaces.

With those definitions, the BHZ tight-binding Hamiltonian on a square lattice 
and in momentum space reads
\be
H_0 = \sum_\bk\,  \Psi^\dagger_\bk  \; H_{\mathrm{BHZ}}(\bk) \;\Psi_\bk 
\label{BHZ-Ham}
\ee
where $\bk$ is defined in the first Brillouin zone,
$\Psi^\dagger_\bk = \big(s^\dagger_{\bk\up},
s^\dagger_{\bk\down},
p^\dagger_{\bk\up},
p^\dagger_{\bk\down}\big)$
are four component spinors and $H_{\mathrm{BHZ}}(\bk)$ is the $4\times 4$ matrix
\beal
H_{\mathrm{BHZ}}(\bk) &= \Big(M- t\,\big(\cos k_x+\cos k_y \big)\Big) \,\sigma_0\otimes\tau_3   \\ 
 &\quad +\lambda\,\sin k_x\;\sigma_3\otimes\tau_1
-\lambda\,\sin k_y\;\sigma_0\otimes\tau_2\, 
\label{eq:H_0}
\eal
Hereafter, we set $t = 1$, the energy unit, $\lambda = 0.4$, half-filled density, i.e., two electrons per site, and, without loss of generality, $M\geq 0$.
For $M<2$ the Hamiltonian \eqn{H_0} describes a QSHI, otherwise a conventional nontopological band insulator (BI). The periodic model \eqn{BHZ-Ham} is invariant under time-reversal $\mathcal{T}$, particle-hole $\mathcal{P}$, inversion $\mathcal{I}$ and $C_4$ symmetries, as well as under spin $U(1)$ rotations around the $z$-axis.

The phase diagram of the model in presence of a Coulomb repulsion between the electrons has attracted a lot of interest in the field \cite{BHZ_AFMTI_DMFT/NRG,BHZ_AFMTI_VCA,Sangiovanni2013, BHZAdriano1, BHZAndrea, BHZAdriano2, BHZCrippa, MottZerosGiorgio, BHZ_AFMTI_DMRG, BHZFrancesca, BHZIvan}. 

The onsite Coulomb interaction projected onto the Wannier orbital basis can be written as
\beal
H_1 &= \sum_\bR\, H_\text{int}(\bR)\,,
\label{interaction}
\eal
where $\bR$ labels the lattice sites and 
\beal
H_\text{int}(\bR) &= 
U_s\,n_{s\bR\up}\,n_{s\bR\down} 
+ U_p\,n_{p\bR\up}\,n_{p\bR\down} \\
&\qquad + V\,n_{s\bR}\,n_{p\bR} + H_\text{dip}(\bR)\,,
\label{Hint}
\eal
with $n_{a\bR\sigma}$ the number of spin $\sigma=\up,\down$ electrons 
at site $\bR$ in orbital $a=s,p$, and 
$n_{a\bR}=n_{a\bR\up}+n_{a\bR\down}$.
The term $H_\text{dip}(\bR)$ in \eqn{Hint} is the 
the dipole component of the multipole expansion of the Coulomb interaction, 
\beal
&H_\text{dip}(\bR) = \fract{J}{2}\,\bigg\{
\Big(\Psi^\dagger_\bR\,\sigma_0\otimes\tau_1\,\Psi^\dagga_\bR\Big)^2\\
&\qquad \qquad \qquad \qquad+\Big(\Psi^\dagger_\bR\,\sigma_3\otimes\tau_2\,\Psi^\dagga_\bR\Big)^2\bigg\}\\
&\qquad = J\,\Big(s^\dagger_{\bR\up}\,s^\dagger_{\bR\down}\,
p^\dagga_{\bR\down}\,p^\dagga_{\bR\up}
+ p^\dagger_{\bR\up}\,p^\dagger_{\bR\down}\,
s^\dagga_{\bR\down}\,s^\dagga_{\bR\up} \\
&\qquad\qquad\quad   + s^\dagger_{\bR\up}\,p^\dagger_{\bR\up}\,s^\dagga_{\bR\up}\,p^\dagga_{\bR\up}
+ s^\dagger_{\bR\down}\,p^\dagger_{\bR\down}\,s^\dagga_{\bR\down}\,p^\dagga_{\bR\down}\Big)\,,
\label{interaction-dipole}
\eal
where $\Psi^\dagga_\bR$ is the Fourier transform in real space of the spinor $\Psi^\dagga_\bk$, 
$\bR$ being a lattice site.
All parameters, $U_s$, $U_p$, $V$ and $J$ 
are positive. The interaction \eqn{Hint} enforces Hund's rules when $\text{min}(U_s,U_p)>V$, which we assume hereafter, and entails that the lowest energy two-electron configuration  
of $H_\text{int}(\bR)$ is the $S_z=\pm 1$ doublet of the  spin triplet configuration. 
For the purpose of this work, in the dipole component $H_\text{dip}(\bR)$ we replace the term $J \, S^{z}_s \, S^{z}_p$ with the $SU(2)$-symmetric form $J \, \vec{S}_s \cdot \vec{S}_p$, in order to recover the more commonly studied Hubbard-Kanamori interaction \cite{Kanamori1963}. This modification mainly affects the atomic limit, where the lowest energy two-electron configuration of $H_\text{int}(\bR)$ becomes the $S=1$ one, thus threefold degenerate. We have also verified that this change does not impact the results discussed in the following sections, therefore we proceed with this choice without loss of generality.

\subsection{The quasiparticle picture in ghost Gutzwiller}
The ghost Gutzwiller (gGut) ansatz~\cite{Lanata2017,guerci2019,frank2021,Mejuto2023a,Lee2023a,Lee2023b,Guerci2023} is a nonperturbative, variational wave function method generalizing the traditional Gutzwiller approach~\cite{Gutzwiller1963,Gutzwiller1965,Bunemann1998,Fabrizio2007,Yao2014,Yao2015,lanata2015,fabrizio2017}.
As such, it proposes a wave function for the full system $\ket{\Psi_\mathrm{G}} = P\ket{\Psi_\mathrm{qp}}$, composed of a 
single Slater determinant $\ket{\Psi_\mathrm{qp}}$ acted upon with a projection operator $P$.
The parameters defining both these ingredients are varied to minimize the ground state energy.
Within an infinite dimensional approximation, this variational optimization can be exactly substituted by a self-consistent embedding of a local impurity Hamiltonian $H_\mathrm{imp}$ within a quasiparticle, hence noninteracting, model of the full system $H_\mathrm{qp}$~\cite{lanata2015}.
This quasiparticle Hamiltonian, of which $\ket{\Psi_\mathrm{qp}}$ is the ground state, can model both low and high energy features of correlation by introducing auxiliary orbitals, the eponymous ghosts.
The inclusion of these ghost orbitals has enabled the accurate description of the Hubbard bands along the Mott transition~\cite{Lanata2017,Lee2023a}, beyond the reach of the traditional Gutzwiller description~\cite{BrinkmanRice}, as well as correlated multi-orbital phenomena~\cite{Mejuto2023a,Lee2023b}, \emph{ab initio} description of covalent bond breaking~\cite{Mejuto2024}, interaction driven magnetic 
phases~\cite{Giuli2025,Bellomia_intracorr}
and spinon excitations in Mott insulators~\cite{Tagliente2025}, to mention some recent applications. 
Crucially, the gGut embedding is formulated in terms of the one-body local reduced density matrix $\langle c^\dagger_{\alpha}c^\dagga_{\beta}\rangle$, resulting in a computationally inexpensive formalism, allowing efficient phase space explorations, and even inspiring hybrid quantum-classical implementations~\cite{Yao2021,Besserve2022,Chen2025,Sriluckshmy2025}.
For details on the derivation and implementation of the gGut method, we refer to the existing literature~\cite{lanata2015,Mejuto2023a,Mejuto2024,Tagliente2025}.
Here we will briefly focus on the main ingredients necessary for extracting topological invariants within the gGut solution, and summarize the equations of the framework in the appendix.
We further note that the Gutzwiller framework as presented here is equivalent to the most common implementations of the rotationally invariant slave-boson approach~\cite{Gebhard2007}.

As mentioned above, the embedding formulation of gGut rests on two complementary model descriptions of the lattice under study: the local, interacting impurity Hamiltonian $H_\mathrm{imp}$ and the extended, noninteracting quasiparticle Hamiltonian $H_\mathrm{qp}$. Upon convergence, the self-consistency condition enforces that the bath one-body density matrix in the impurity model $\Delta^{\rm imp}_{ab}\equiv\langle d_b d_a^\dagger\rangle_{\rm imp}$ equals the local quasiparticle one-body density matrix $\Delta^{\rm qp}_{ab}\equiv\langle d_{0a}^\dagger d_{0b}\rangle_{\rm qp}$ (see Appendix A, Eq.\,\ref{eq:SI_scf})
Through this feedback loop, $H_\mathrm{qp}$ can capture, upon introduction of the ghost orbitals, the local correlated effects present in the interacting $H_\mathrm{imp}$ at low and high energy within a band structure picture.
For a periodic system, $H_\mathrm{qp}$ follows~\cite{Mejuto2023a}
\beal
    H_\mathrm{qp}(\bk) = \sum_{\bk,ab}\left(\sum_{\alpha\beta}R^\dagger_{a\alpha}\epsilon^{\alpha\beta}_{\bk}R^\dagga_{\beta b}-\lambda_{ab}\right)d^\dagger_{\bk,a}d^\dagga_{\bk,b},
    \label{eq:Hqp}
\eal
where $\alpha, \beta$ run over the physical orbitals in the original system, including spin, and $\epsilon^{\alpha,\beta}_{\bk}$ represents the original dispersion relation of the lattice.
Meanwhile, $a, b$ run over the quasiparticle orbitals, which include the ghosts, and we have introduced the renormalization factors $R_{\alpha a}$ and the local, effective one-body potential $\lambda_{ab}$.
Essentially, $H_\mathrm{qp}$ carries over the dispersion relation of the underlying, noninteracting lattice, and recovers correlation effects by renormalizing the hopping terms through $R$, introducing a local effective one-body interaction $\lambda$. Crucially, the accuracy of the account of correlations is greatly enhanced by enlarging the Hilbert space with the addition of ghost orbitals, particularly with regards to high energy spectral features.
It is this representation in $H_\mathrm{qp}$ that realizes an effective band structure theory for correlated electrons, which we leverage to determine topological invariants.

\subsection{Spectral representations of correlated electrons}
\label{sec:SpectralRep}
Before discussing how topological invariants can be accessed from $H_\mathrm{qp}$, some words on the spectral representation of strongly correlated electrons are in order.
The one-body spectrum of a many-body electron system, i.e.\,the charged, one-electron excitations it supports, are contained within the one-body Green's function $G(\omega,\bk)$,with $\omega$ a real frequency.
More generally, $G(z,\bk)$ is a matrix-valued function of crystal-momentum $\bk$ and complex-frequency $z$. In this section we will work on the Matsubara axis, where $z =i\epsilon$ is an imaginary Matsubara frequency.
In a noninteracting system, where the Hamiltonian $H_0(\bk)$ is itself a one-body operator, the Green's function can be simply evaluated as the resolvent
\beal
    G_0(i\epsilon,\bk) = \frac{1}{i\epsilon - H_0(\bk)}.
    \label{eq:G0}
\eal
For an interacting system, where $H$ is a many-body operator, it is necessary to project the resolvent into the one-electron space, usually in terms of expectation values over a single pair of creation/annihilation operator such as $\langle c^\dagger_\alpha\ \frac{1}{i\epsilon-H}\ c^\dagga_\beta\rangle$.
Alternatively, one can express the one-body Green's function of an interacting system in terms of a simple resolvent by introducing the self-energy $\Sigma(i\epsilon,\bk)$.
This is essentially the effective, frequency-dependent one-body potential that needs to be added to the noninteracting part of the Hamiltonian to recover electronic correlation.
Once the one-body Green's function is written in terms of the self-energy, a few manipulations can be introduced to make the relation between a general self-energy and Fermi-liquid theory more apparent~\cite{Fabrizio2022,Fabrizio2023,AndreaTopoZeros}
\beal
G(i\epsilon, \mathbf{k}) 
& = \frac{1}{i\epsilon - H_{0}(\bk) - \Sigma(i\epsilon, \bk) } \\
& = \sqrt{Z(\epsilon, \bk)} \, \frac{1}{i\epsilon - H_{*}(\epsilon,\bk) \;}\,
 \sqrt{Z(\epsilon, \bk)} \\
& \equiv \sqrt{Z(\epsilon, \bk)} \,   G_{*}(i\epsilon, \bk) \,\sqrt{Z(\epsilon, \bk)}
\label{G*}
\eal
where, we have introduced the semi positive-definite matrix 
\be
Z( \epsilon, \mathbf{k} ) = Z( -\epsilon, \mathbf{k} )=\bigg( 1 - \frac{ \Sigma_2( i\epsilon, \mathbf{k} ) }{\epsilon} \bigg)^{-1}
\label{Z}
\ee
in terms of
\beal
\Sigma_1( i\epsilon, \mathbf{k} ) & = \frac{\Sigma(i\epsilon, \mathbf{k} ) + \Sigma(i\epsilon, \mathbf{k} )^\dagger}{2}\;,\\
\Sigma_2( i\epsilon, \mathbf{k} ) & = \frac{\Sigma(i\epsilon, \mathbf{k} ) - \Sigma(i\epsilon, \mathbf{k} )^\dagger}{2i}\;, 
\eal
which are themselves Hermitian since $\Sigma(i\epsilon,\bk)^\dagger=\Sigma(-i\epsilon,\bk)$.
The effective Hamiltonian $H_{*}(\epsilon,\bk)$ introduced in Eq.\,\eqref{G*}, and which follows 
\beal
&H_{*}(\epsilon, \mathbf{k} ) = H_{*}(-\epsilon, \mathbf{k} )\\
&\quad =\sqrt{Z(\epsilon, \bk)}\, \big( H_{0}(\mathbf{k}) + \Sigma_1( i\epsilon, \mathbf{k} ) \big)\, \sqrt{Z(\epsilon, \bk)}\,,
\label{H*}
\eal
is therefore also sometimes referred to as quasiparticle Hamiltonian within the scope of Fermi-liquid theory.
As we will discuss in the following section,
$H_*(\bk)=H_{*}(\epsilon\to 0,\bk)$
is commonly employed in the study of strongly correlated systems as a renormalized one-body effective Hamiltonian to characterize the integrated topology encoded in the Green's function winding number~\cite{Ishikawa1, Ishikawa2, AndreaTopoZeros, PasquaSciPost}. 
To avoid confusion with $H_\mathrm{qp}$ in the gGut ansatz, here we will use the term Fermi-liquid quasiparticle Hamiltonian for $H_{*}$.

Now, within the gGut approximation we can also access the one-body Green's function of the lattice system.
The usual recipe involves taking the resolvent of the quasiparticle Hamiltonian $H_\mathrm{qp}$ and projecting it back to the physical orbital space with $R$ following
\beal
    G^{\mathrm{gGut}}_{\alpha\beta}(i\epsilon,\bk) = \sum_{ab}R_{\alpha a}^\dagga\left[\frac{1}{i\epsilon-H_\mathrm{qp}(\bk)}\right]_{ab}R^\dagger_{b\beta}.
    \label{eq:GgGut}
\eal
In essence, we build the physical Green's function as a linear combination of the noninteracting, quasiparticle bands.
It is in this sense that gGut provides a band structure theory for correlated electrons, and which we will leverage to evaluate topological invariants in the following subsection.
While \emph{a priori}, and despite its shown success~\cite{Lu2013,Nandy2024,Lanata2017,guerci2019,Mejuto2023a,Lee2023a,Lee2023b,Mejuto2024,Giuli2025,Tagliente2025}, this prescription may seem somewhat \emph{ad hoc}, we argue that it has the right mathematical form to potentially recover the exact one-body Green's function in Eq.\,\eqref{G*}. We conclude remarking that Eq.\,\ref{eq:GgGut} represents the Green's function spectrum as a finite pole expansion. In this sense, gGut captures the positions and weights of both coherent and high-energy features, but cannot account for the lifetime decay and the broadening of excitations away from the Fermi level. Indeed, one can always split the positive semi-definite matrix $Z(\epsilon,\bk)$ as
\beal
    Z(\epsilon,\bk) = A(\epsilon,\bk)\,A^\dagger(\epsilon,\bk)\,,
    \label{eq:ZintoA}
\eal
where in principle $A(\epsilon,\bk)$ can be a rectangular matrix with a number of rows $N$ equal to the number of physical orbitals, but a larger number of columns $M$.
With this, the Fermi-liquid quasiparticle Hamiltonian $H_{*}$ in Eq.\,\eqref{H*} follows
\be
H_*(\epsilon,\bk) \to A^\dagger(\epsilon,\bk)\,
\big( H_{0}(\mathbf{k}) + \Sigma_1( i\epsilon, \mathbf{k} ) \big)\,A(\epsilon,\bk)\,,
\label{eq:H*2}
\ee
becoming an $M\times M$ Hermitian matrix.
Consequently, the physical, interacting one-body Green's function can also be rewritten in terms of the rectangular $A(\epsilon,\bk)$ matrices as
\beal
G(i\epsilon,\bk)=A(\epsilon,\bk)\;
\fract{1}{\;i\epsilon - H_*(\epsilon,\bk)\;}\;
A^\dagger(\epsilon,\bk)\,.
\label{eq:GwithA}
\eal
At first glance, there seems to be no point in choosing $M > N$, hence a number of quasiparticles larger than the number of physical orbitals.
However, inspecting Eq.\,\eqref{eq:GwithA}, it becomes apparent that such a strategy can lead to an important simplification: if $M$ is large enough, it should become possible to approximate the full $G(i\epsilon,\bk)$ with frequency-independent $A(\bk)$ and $H_{*}(\bk)$, namely
\beal
G(i\epsilon,\bk) \simeq A(\bk)\;
\frac{1}{i\epsilon - H_*(\bk)}\;
A^\dagger(\bk)\,.
\label{eq:GtoGqp}
\eal
After all, for $M$ going to infinity, this is always true as can be seen inspecting the Lehman representation of the Green's function
\beal
G_{\alpha \beta}(i \epsilon, \bk ) = & \sum_n \frac{\bra{n} c_{\alpha \bk} \ket{0} \bra{0} c^\dagger_{\beta \bk} \ket{n}}{ i \epsilon - ( E_0 - E_n )} \\
& + \sum_n \frac{\bra{0} c_{\alpha \bk} \ket{n} \bra{n} c^\dagger_{\beta \bk} \ket{0}}{ i \epsilon - ( E_n - E_0 )}
\eal
where $\alpha, \, \beta$ label the physical orbitals as before and $n$ the many-body eigenstates. If we now define
\begin{equation}
A_{\alpha n}(\bk) = 
\begin{cases}
\bra{n} c_{\alpha \bk} \ket{0} & \text{if } \epsilon_n (\bk) = E_0 - E_n < 0, \\
\bra{0} c_{\alpha \bk} \ket{n} & \text{if } \epsilon_n (\bk) = E_n - E_0 > 0,
\end{cases}
\end{equation}
then the Green's function can be written as
\beal
G_{\alpha \beta}(i \epsilon, \bk ) = \sum_n A_{\alpha n}(\bk) \frac{1}{i \epsilon - \epsilon_n (\bk)} A^{*}_{\beta n}(\bk )
\eal
where the number of columns of $A_{\alpha n}(\bk)$ grows exponentially with the system size, as the number of many-body states labeled by $n$.

Identifying $A(\bk)\equiv R(\bk)$, we almost arrive at the gGut approximation for the Green's function in Eq.\,\eqref{eq:GgGut}.
The only missing ingredient is to remove the formal momentum dependence of the projection matrices in Eq.\,\eqref{eq:GtoGqp}.
This can be interpreted as a local, hence $\bk$-independent, approximation, which becomes natural considering the infinite dimensional limit implicit in the embedding formulation of gGut.
Thus, the gGut approximation to the Green's function in Eq.\,\eqref{eq:GgGut} is well justified as a finite truncation of the Lehman expansion to the exact Green's function within a local embedding approximation.
Armed with this, we can tackle the extraction of topological invariants from this band structure representation of strongly correlated electrons.

We note that pole-expansion based expressions have been explored in the context of correlated topology~\cite{Savrasov2006,XiDai_poleH,Ivanov2019,Held_poleH}. 
These typically are constructed as few-pole approximations to an otherwise computed self-energy, e.g.\,through DMFT. In contrast, we stress that the pole-expansion of the Green's function in gGut, and thus the quasiparticle band description, is naturally built into the ansatz:
the position and bandwidths of the ghost bands are fixed through the embedding approximation to the variational principle.

\subsection{Topological invariants within ghost Gutzwiller}
We are finally ready to examine how the band-structure emerging from $H_\mathrm{qp}(\bk)$ can be used to extract the topological invariants of the system. Specifically, following the previous discussion on spectral representations of correlated electrons, we demonstrate the physical relevance of computing Chern numbers for these quasiparticle bands, and their relation to the Chern number of the fully interacting system.

To this end, we note that the gGut Green's function introduced in Eq.\,\eqref{eq:GgGut} embodies the notion that the quasiparticle Hamiltonian, although optimized to describe ground state properties, also provides a meaningful approximation to single-particle excitations.
This perspective becomes even more compelling when considering the robust quantization of topological invariants of the ground state. The energy- and momentum-resolved features of the approximate single-particle excitations in $H_\text{qp}(\bk)$ are expected to be protected by the intrinsic robustness of the topological invariants they support.
In this sense, it is reasonable to expect that $H_\text{qp}(\bk)$ in Eq.\,\eqref{eq:Hqp} should properly describe the topological features of the original, interacting system.
It is instructive to work out this description in a particular example, and thus the rest of this subsection is devoted to explicitly establishing the connection between the gGut quasiparticle Hamiltonian and the winding number of the physical Green's function, a central quantity in the description of ground state topology in self-energy based methods like DMFT.

The single-particle Green's function winding number can be written as~\cite{Ishikawa1, Ishikawa2, AndreaTopoZeros}
\beal
&W[G] =  \frac{1}{24 \pi^{2}}\!\int\!\! \mathrm{d}\epsilon \, \mathrm{d}\bk \, \epsilon_{\mu \nu \rho} \text{Tr} \Big( G(i\epsilon, \bk) \, \partial_{\mu} G(i\epsilon, \bk)^{-1} \, \\
& \qquad\; G(i\epsilon, \bk) \, \partial_{\nu} G(i\epsilon, \bk)^{-1} \, G(i\epsilon, \bk) \, \partial_{\rho} G(i\epsilon, \bk)^{-1} \, \Big ),
\label{WindingNumber}
\eal
where $G(i\epsilon, \bk)$ is the interacting Green's function in the Matsubara frequency domain. When perturbation theory holds, $W[G]$ coincides with the zero-temperature Hall conductance in units of $e^2 / 2 \pi \hbar$ \cite{Ishikawa1, Ishikawa2, AndreaTopoZeros, PasquaSciPost}. This makes the winding number a particularly powerful tool for methods that grant access to the full single-particle Green's function, such as DMFT and gGut. However, the direct evaluation of \eqn{WindingNumber} is often computationally expensive. A common alternative is to exploit the fact that this winding number is equivalent to the Chern number of the Hermitian matrix $-G(0, \bk)^{-1}$ \cite{ZhangWinding}, or equivalently $-G_{*}(0, \bk)^{-1}$ \cite{AndreaTopoZeros}, which is related to the Fermi-liquid quasiparticle Hamiltonian in Eq.\,\eqref{H*}.

We mentioned that within the gGut framework the physical single-particle thermal Green's function $G(i\ep,\bk)$, which is an $N\times N$ matrix, is approximated by 
\beal
G(i\ep,\bk) \approx G^\text{gGut}(i\ep,\bk) =R\,G_\text{qp}(i\ep,\bk)\,R^\dagger\,,
\label{G vs Gqp}
\eal
where 
\bealn
G_\text{qp}(i\ep,\bk)= \fract{1}{\;i\ep-
H_\text{qp}(\bk)\;}\;,
\eal
is the $M\times M$ quasiparticle Green's function matrix with $M\geq N$. 
In \eqn{G vs Gqp}, $R$ is an $N\times M$ matrix that we assume semi unitary: $R\,R^\dagger=\mathbb{I}_N$, the $N\times N$ identity, while $R^\dagger\,R=P$ a projector, 
$P^2=P$. This assumption can be always enforced upon rescaling 
\bealn
R \to \fract{1}{\;\sqrt{R\,R^\dagger\,}\;}\;R\;.
\eal
Through the definition \eqn{WindingNumber}, we find that 
\beal
W[G] &\approx W\left[G^\text{gGut}\right] 
= W\left[P\,G_\text{qp}\,P\right]
\,.
\label{eq:TopoConj-1}
\eal
However, since the inverse of $P\,G_\text{qp}\,P$ does not simply equal $P\,G_\text{qp}^{-1}\,P$, calculating $W\left[P\,G_\text{qp}\,P\right]$ directly is not  straightforward. Nonetheless, one would expect that the non-trivial topological character of $G_\text{qp}$ is solely due to the projected physical states,
namely that     
\beal
W\left[P\,G_\text{qp}\,P\right] \approx 
W\left[G_\text{qp}\right]\,,
\label{eq:TopoConj}
\eal
especially if the edge modes of $H_\text{qp}(\bk)$ appear also in the spectrum of $G^\text{gGut}(i\ep,\bk)$. 
Then calculating $W\left[G^\text{gGut}\right]$ does not require the explicit computation of the Green's function: one merely requires to evaluate the Chern number of the quasiparticle bands. 

We conclude by stressing that $H_\mathrm{qp}(\bk)$ can provide even more topological information than merely the winding number \eqn{WindingNumber}. The latter is, in fact, an integrated quantity that does not convey information about the energy and momentum resolved topological features useful for interpreting experimental spectra. Conversely, the quasiparticle Hamiltonian $H_\mathrm{qp}(\bk)$ in gGut offers this information as well without any additional computational cost. This feature grants access to a band structure that, unlike DFT, also encompasses the Hubbard bands, and thus enables a direct comparison with experimental data under the assumption that the quasiparticle bands accurately represent the genuine single-particle spectrum. 

\begin{figure}[t!]
    \centering
    \includegraphics[width=0.47\textwidth]{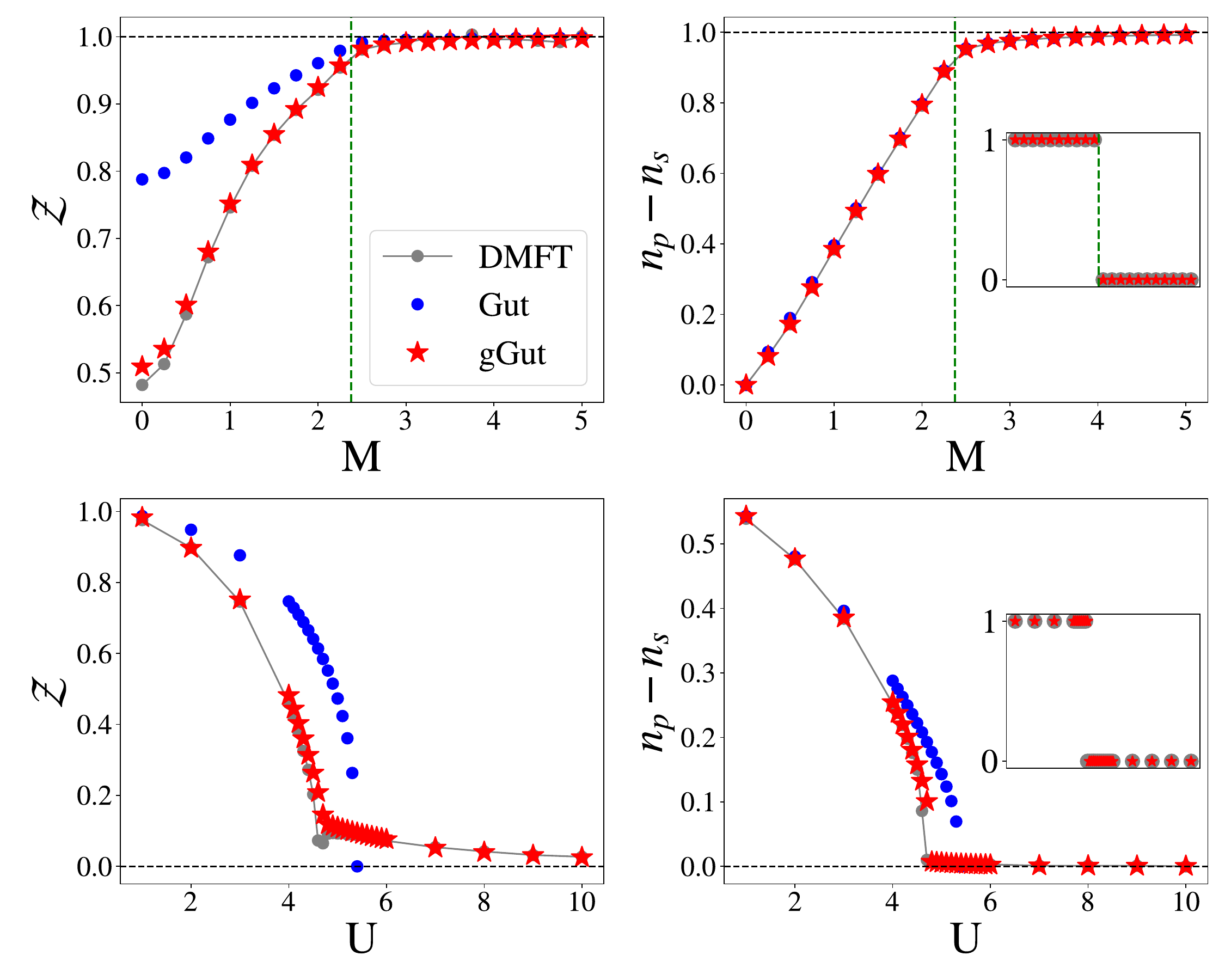}
    \caption{Comparison between different methods on two representative cuts of the phase diagram. Upper panels: quasi-particle residue $\mathcal{Z}$, related to the effective-mass enhancement $m^*/m \sim 1/\mathcal{Z}$}, and on-site orbital polarization $n_p - n_s$ at fixed $U=3.0$, as a function of the bare mass $M$. The inset in the right plot shows the quantized topological invariants computed withing gGut (red stars) and DMFT (gray dots). The green dashed line marks the topological transition between the QSHI and the BI phases. Lower panels: same analysis at fixed $M=1.0$, across the transition from the QSHI to the MI as a function of the interaction strength $U$.
    \label{fig:Benchmarks}
\end{figure}

\section{Bulk BHZ Model: Quasiparticle Perspective and DMFT Comparison}
In this section we use the interacting BHZ model \ref{sec:Interacting BHZ model} to validate the gGut quasiparticle-band 
framework in a setting where band topology and local correlations are both essential. This model has been extensively studied 
with DMFT and extensions~\cite{Sangiovanni2013, BHZAdriano1, BHZAdriano2, BHZCrippa, MottZerosGiorgio, BHZFrancesca, BHZIvan}, 
providing established reference results for both correlation indicators (effective mass renormalization) and topological indices. 
Here we focus on two standard correlation-driven effects in the phase diagram: the renormalized QSHI-BI transition and the 
QSHI-MI transition.  To enable a direct, parameter-matched comparison,
 we complement the existing literature with our own single-site DMFT calculations performed for the same Hamiltonian parameters
used in the gGut study. The excellent agreement with DMFT, itself a well-established non-perturbative method 
for strongly correlated systems, validates the accuracy of the gGut approach across all coupling strengths. For the DMFT calculations, we used the impurity solver implemented in the TRIQS library \cite{TRIQS}, employing a continuous time quantum Monte Carlo algorithm based on a hybridization expansion of the partition function \cite{CTQMC1,CTQMC2}. The DMFT calculations were obtained at $\beta = 50$, in units of the hopping amplitude $t$ (see Eq.\,\ref{eq:H_0}), and the results do not change lowering the temperature. The gGut data were obtained for $T = 0$. 

The interaction parameters of \eqn{Hint} chosen for this section are: $U_s = U_p \equiv U$, $V= U - 2J$ and $J = U / 4$, following the standard parametrization adopted in previous DMFT studies of the interacting BHZ model to enable a direct benchmark ~\cite{Sangiovanni2013, BHZAdriano1, BHZAdriano2, BHZCrippa, MottZerosGiorgio, BHZFrancesca, BHZIvan}. In the weakly correlated regime the primary role of the interaction is to shift the topological transition between the QSHI and the BI towards higher values of the on-site bare mass $M$. 
Within the gGut framework the QSHI–BI transition is signaled by a closure and reopening of the gap of the quasiparticle Hamiltonian $H_{\mathrm{qp}}(\mathbf{k})$, i.e.\,by a band inversion at the critical value of $M_c$. In the parameter range considered here, the gap closes via a linear (Dirac) band crossing of the quasiparticle bands. Accordingly, the QSHI–BI boundary discussed in Fig.\,\ref{fig:Benchmarks} corresponds to a continuous topological band-inversion transition \cite{Chen2019}.
As we can see in Fig.\,\ref{fig:Benchmarks} the transition, for fixed $U=3$, appears at $M_\mathrm{c} \simeq 2.3$. Already the Gutzwiller (Gut) solution without ghosts can accurately capture the behavior of the on-site orbital polarization $n_p-n_s$, nonetheless it severely underestimates the role of the correlations, e.g., in the quasiparticle renormalization matrix $Z$. Nonetheless, just by including two ghosts per orbital in the quasiparticle Hamiltonian $H_\mathrm{qp}(\bk)$, the gGut solution can accurately describe correlation effects, matching the DMFT predictions. These auxiliary degrees of freedom correctly capture charge fluctuations in the systems, allowing for the developing of high-energy Hubbard bands. We also note that the quasiparticle renormalization matrix
\be
Z \equiv \bigg ( 1 - \frac{\partial \text{Re}\Sigma(\omega) }{\partial \omega}\bigg|_{\omega \to 0} \bigg )^{-1} ,
\ee
where $\omega$ is a real frequency, is diagonal thanks to the symmetries of the local Green's function and self-energy: $Z = \mathcal{Z}\otimes \mathbb{I}_{4 \times 4}$, where $\mathcal{Z}$ is related to the effective-mass enhancement $m^*/m \sim 1/\mathcal{Z}$. In the inset of the right plots of Fig.\,\ref{fig:Benchmarks}, we see perfect agreement between the topological invariant computed from $H_\mathrm{qp}(\bk)$ and the winding number of the Green's function obtained in DMFT.
\begin{figure}[t!]
    \centering
    \includegraphics[width=0.5\textwidth]{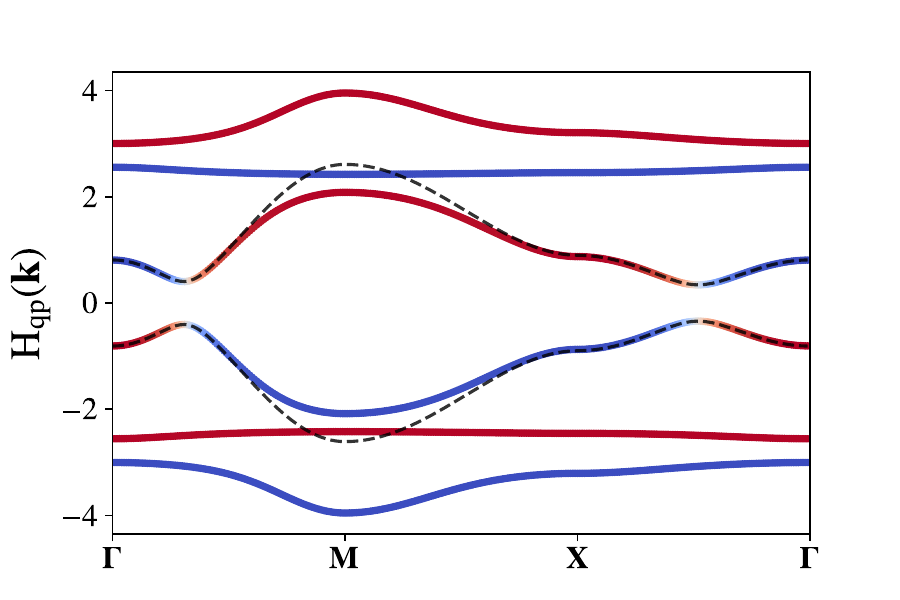}
    \caption{Band-structure of the quasiparticle Hamiltonian $H_{\mathrm{qp}}(\mathbf{k})$ along a high-symmetry path in the Brillouin zone. The color scale indicates the orbital character of the eigenstates: red for $s$-like and blue for $p$-like contributions. The two energy bands close to the Fermi level have a mixed-orbital character and are topologically nontrivial. In contrast, the high-energy bands, associated with the dispersive Hubbard bands, are topologically trivial. For comparison, black dashed lines show the dispersion of $H_{*}(\mathbf{k}) = \sqrt{Z} \left ( H_{0}(\mathbf{k}) + \text{Re} \Sigma(0, \mathbf{k} )  \right ) \sqrt{Z}$, which encodes the topological content of the Green's function winding number. }
    \label{fig:QPHamiltonian}
\end{figure}

Focusing now on a cut of the phase diagram at fixed $M$ and varying $U$ we can drive the system towards a strongly correlated Mott insulator (MI). The Gut solution correctly describes the qualitative renormalization of low-energy bands of the QSHI before the onset of a MI, but it fails to faithfully describe the Mott insulating phase. This is a consequence of what is observed in the Mott transition in a single-band Hubbard model at half-filling \cite{BrinkmanRice}: the Gutzwiller ansatz satisfactorily accounts for the narrowing and eventual vanishing of the coherent metallic band at the Fermi level, but it fails to describe the incoherent high-energy Hubbard bands, which characterize the nature of the insulating state and relate it to the atomic limit. Only the inclusion of the ghosts allows for a meaningful description of the MI, which perfectly agree with the DMFT solution as shown in the bottom panels of Fig.\,\ref{fig:Benchmarks}. In the predicted MI phase the integrated topology encoded in the winding number is trivial, consistently with the fact that the zeros of the Green's function are flat and trivial \cite{AndreaTopoZeros} in any local approximations.

We conclude this section with a closer inspection to the band-structure arising in gGut from the quasiparticle Hamiltonian $H_{\mathrm{qp}}(\bk)$ of the auxiliary fermionic degrees of freedom. In Fig.\,\ref{fig:QPHamiltonian}, we show the eigenvalues of $H_{\mathrm{qp}}(\bk)$ for the correlated QSHI at $M = 1.5$ and $U=3.0$. The color code reflects the $s$- and $p$-orbital character of the corresponding eigenstates. The two energy bands close to the Fermi level are topologically nontrivial, as suggested by their mixed-orbital character and confirmed from the calculation of the spin-resolved Chern number. In contrast, the high-energy bands, associated with the dispersive Hubbard bands, are topologically trivial in this case, but a more detailed discussion about their topological character and physical implications is deferred to section \ref{sec:TopoHubbardBands}. For comparison, we also show the eigenvalues of $H_{*}(\bk)$ (dashed black lines), which encodes the topological content of the Green's function winding number in Eq.\,\ref{WindingNumber}. As shown in the figure, the low-energy bands of $H_{\mathrm{qp}}(\bk)$ closely match those of $H_{*}(\bk)$, particularly near the avoided crossing where the Berry curvature is concentrated. When perturbation theory is valid, both approaches yield consistent information about the topological nature of the system as encoded by the Green's function winding number. However, while $H_{*}(\bk)$ is intrinsically tied to the winding number description, $H_{\mathrm{qp}}(\bk)$ also captures high-energy excitations, including the Hubbard bands and their possible topological features. This establishes $H_{\mathrm{qp}}(\bk)$ as a general and physically grounded tool for analyzing the topology of excited states throughout the spectrum of strongly correlated systems, complementing the successful topological Hamiltonian approach.
\begin{figure}[t!]
    \centering
    \includegraphics[width=0.48\textwidth]{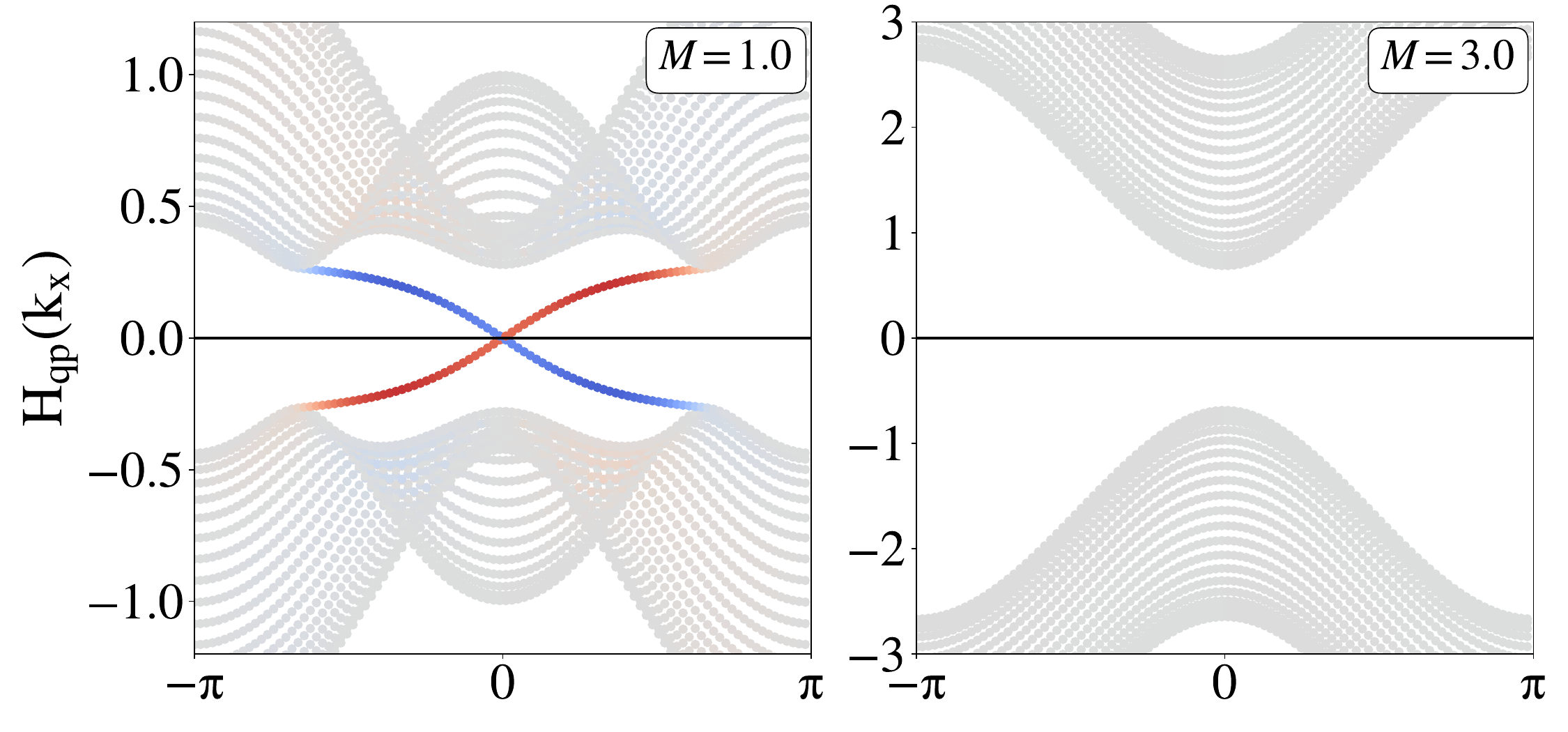}
    \caption{Quasiparticle spectrum of $H_{\mathrm{qp}}(k_x)$ for a slab geometry with OBC along $\hat{y}$ ($N_y = 20$) and PBC along $\hat{x}$. We show the data for two representative points of the phase diagram cut discussed in the upper panels of Fig.\,\ref{fig:Benchmarks} ($U=3.0$). The color code quantifies the degree of localization of the spin-up eigenstates at the upper (blue) and lower (red) edges (for the spin-down the colors are reversed, as expected for helical edge states). In the topological phase, edge states emerge within the bulk gap and cross the chemical potential, consistent with the nontrivial Chern number of $H_\mathrm{qp}(\bk)$ and the bulk-boundary correspondence.}
    \label{fig:OBC_QPHamiltonian}
\end{figure}

\begin{figure*}[t!]
    \centering
    \includegraphics[width=0.94\textwidth]{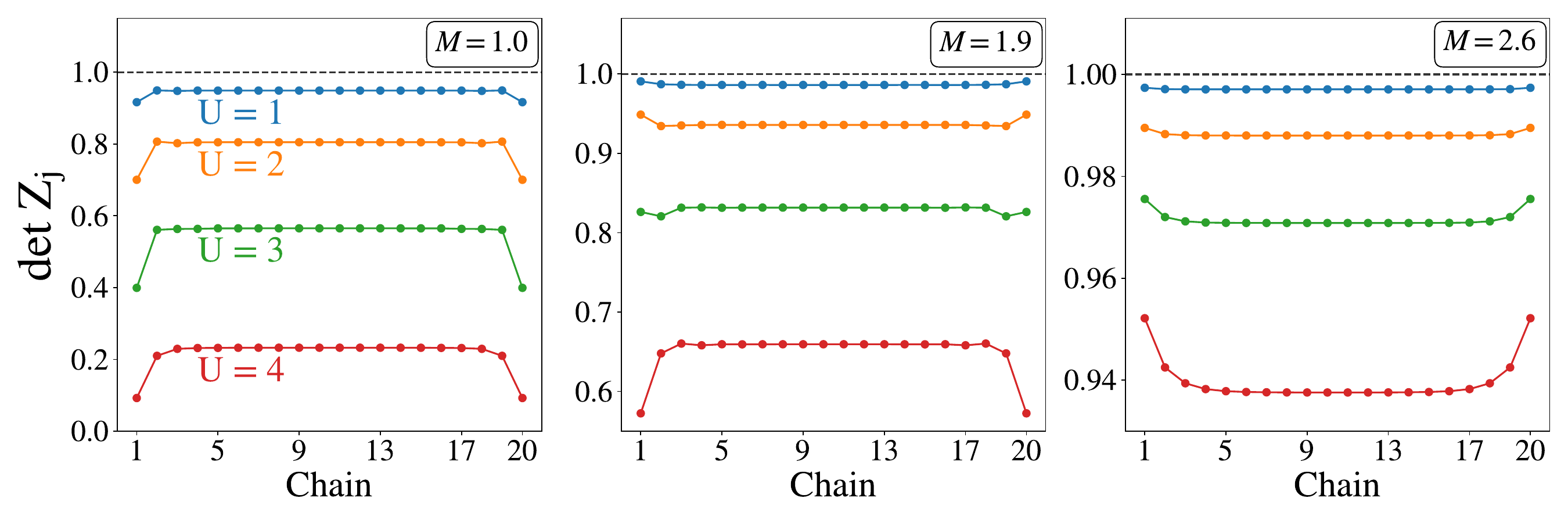}
    \caption{Determinant of the quasiparticle residue matrices $Z_j$ for each chain $j$ for a slab geometry with OBC along $\hat{y}$ ($N_y = 20$) and PBC along $\hat{x}$. Each panel shows results for four values of the interaction strength $U$, at fixed $M$. At the slab boundaries, the reduced hopping due to the interface with the vacuum modifies the local correlations: depending on whether the dominant effect is the effective enhancement of the interaction strength or the increase in the effective mass-to-hopping ratio $M_{\mathrm{eff}}/t_{\mathrm{eff}}$, correlations may be either enhanced or suppressed. In the latter case, the boundaries becomes effectively ``less topological'' and thus less correlated by the interaction. }
    \label{fig:OBC_detZ}
\end{figure*}

\section{OBC calculations and edge states in the quasiparticle Hamiltonian}
An equivalent and more direct way to characterize the band topology of a system is looking for symmetry-protected edge states under open boundary conditions (OBC) along one direction. According to the bulk-edge correspondence \cite{BulkBoundary1, BulkBoundary2}, gapless edge states must emerge at the interface between regions belonging to different topological phases, i.e.\,characterized by distinct topological invariants. However, solving the full many-body problem on a large, finite-size system is numerically prohibitive. A computationally efficient alternative is to study a two-dimensional slab geometry decomposing it as a set of coupled one-dimensional chains. Inter-chain couplings are fully retained in the slab Hamiltonian, while electronic correlations are treated at the single-impurity (local) level with parameters that may depend on the chain index $y$. Consequently, one impurity problem is solved for each inequivalent chain, and the total computational cost scales linearly with the number of chains $N_y$, assuming the impurity solver is the dominant bottleneck. Although the impurity problems are solved separately, they are coupled through the embedding, which is built from the full slab quasiparticle Hamiltonian $H_{\rm qp}$, including inter-chain hopping. Crucially, the impurity problem arising in the gGut embedding scheme is much cheaper compared to approaches like DMFT, enabling the exploration of larger system sizes. This opens the door to studying the interplay between correlations, topology, and finite-size effects in regimes that have mostly been explored in noninteracting systems so-far \cite{ACook1, STraverso}. 

In Fig.\,\ref{fig:OBC_QPHamiltonian} we show the quasiparticle spectrum of $H_{\mathrm{qp}}(k_x)$ for a slab geometry with open boundaries along $\hat{y}$ ($N_y = 20$) and periodic boundary conditions (PBC) along $\hat{x}$. The calculations are performed for two values of $M$, corresponding to representative points along a fixed $U=3.0$ cut of the phase diagram (see Fig.\,\ref{fig:Benchmarks}). For each value of $k_x$, there are $N_\mathrm{orbitals} \times N_\mathrm{ghosts} \times N_y $ eigenvalues and eigenstates per spin. In the topologically nontrivial phase, we observe two helical edge states crossing the chemical potential within the bulk gap. These edge states are exponentially localized at the boundaries, as highlighted by the color coding. Their finite spectral weight in the physical density of states further supports the interpretation of the Chern numbers computed from the occupated bands of $H_{\mathrm{qp}}(\bk)$ as a genuine topological invariant of the interacting system. 

\begin{figure*}
    \centering
    \includegraphics[width=.95\textwidth]{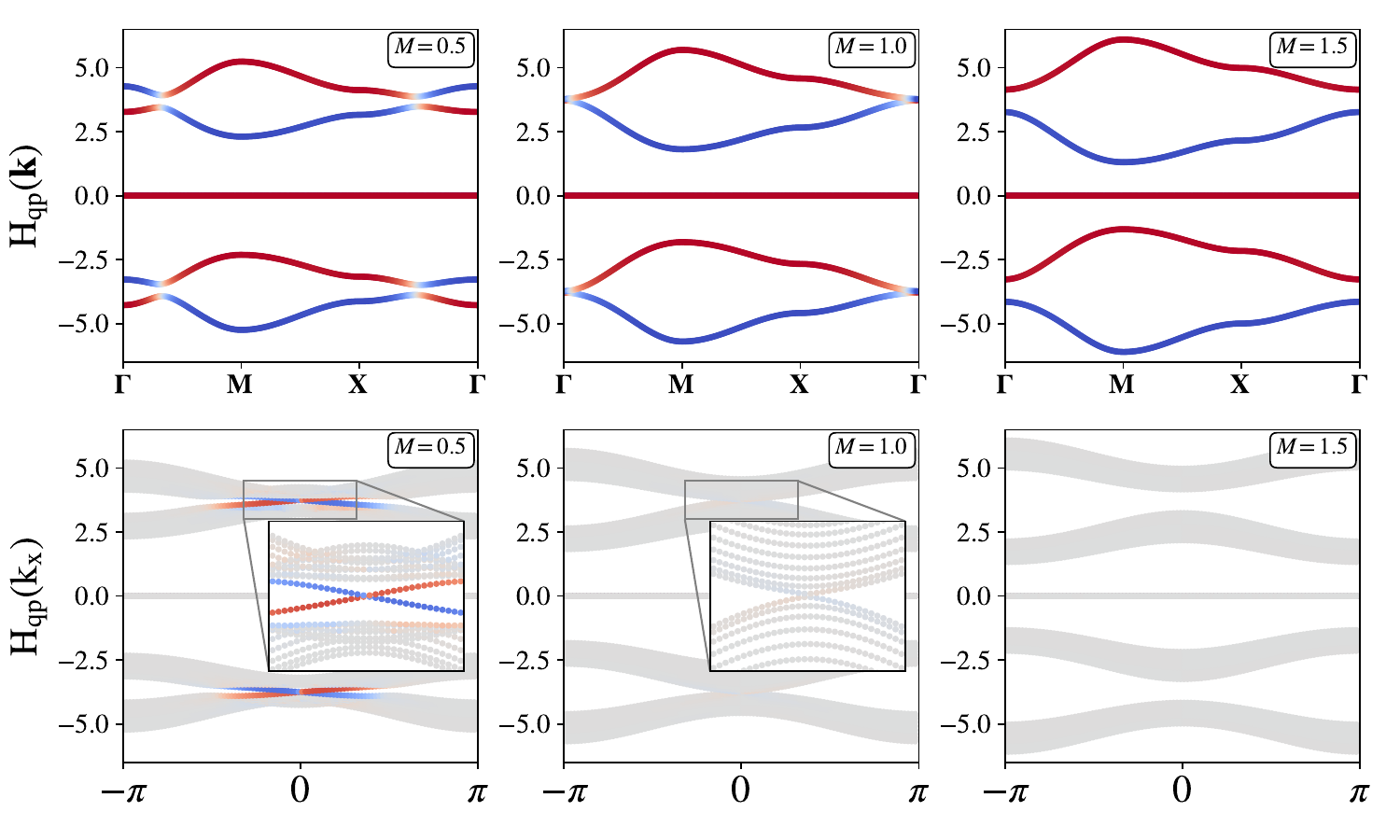}
    \caption{Quasiparticle band structure of $H_\mathrm{qp}(\bk)$ showing the topological Hubbard bands. Upper panels: PBC band structures at $U = 8.0$ and for three values of $M$. The color scale indicates the orbital polarization (red for $s$ and blue for $p$ orbital), highlighting the hybridization and band inversion between the Hubbard bands. Lower panels: spin-up eigenvalues of $H_\mathrm{qp}(k_x)$ in a slab geometry with OBC along $\hat{y}$ ($N_y = 20$). The color code quantifies the degree of localization of the spin-up eigenstates at the upper (blue) and lower (red) edges (for the spin-down the colors are reversed), revealing the emergence of topologically protected helical edge states inside the Hubbard band gaps for $M \lesssim 1$.}
    \label{fig:TopologyHubbardBands}
\end{figure*}

By extending this slab analysis to multiple fixed $M$ cuts of the phase diagram, we can observe a nontrivial and to the best of our knowledge never explored manifestation of the interplay between topology, interactions and boundary effects. Before proceeding, it is important to note that in OBC geometry, the wavefunction renormalization matrix $Z_j$, with $j$ indexing the chains, is no longer diagonal in the orbital basis. This is due to the absence of the symmetry constraints that, under PBC, enforce both the local Green's function and the self-energy to be orbital-diagonal. These off-diagonal elements become significant near the boundaries of a strongly correlated QSHI, but decay rapidly within the bulk. For this reason, in Fig.\,\ref{fig:OBC_detZ}, we plot the determinant of the quasi-particle residue matrix $Z_j$, rather than only the diagonal elements as in the PBC case. We emphasize, however, that both the determinant and the diagonal entries exhibit the same chain-dependent behavior, which we now discuss. 

In the first panel of Fig.\,\ref{fig:OBC_detZ}, we show the spatial profile of $\det{Z_j}$ across the slab at fixed $M=1.0$, as a function of the chain index $j$, for several values of the interaction strength $U$. The boundary chains exhibit stronger renormalization (smaller $Z_j$) than the bulk, indicating enhanced correlations. This is expected, as the reduced kinetic energy at the edges, due to missing neighbors, leads to a relatively stronger effect of the interaction. Interestingly, for $M=1.9$, we find the opposite behavior at small $U$: correlations are suppressed at the boundaries. This counterintuitive result arises because the reduced hopping at the edges increases the effective ratio $M_{\mathrm{eff}} / t_{\mathrm{eff}}$, thereby favoring a topologically trivial phase. As a result, the boundaries are effectively ``less topological'' and hence less correlated by the interaction. Only at larger $U$ does the usual trend of enhanced edge correlations reemerge. Finally, the third panel of Fig.\,\ref{fig:OBC_detZ} shows the spatial dependence of $\det{Z_j}$ for $M = 2.6$, which further support our argument: here the dominant contribution comes from the renormalization of $M_{\mathrm{eff}} / t_{\mathrm{eff}}$ at the edges, therefore the edge renormalization is reversed compared to the $M = 1.0$ case. These results clearly demonstrate how the competition between topological and correlation effects leads to nontrivial boundary behavior in interacting topological insulators.

\section{Topology of the Hubbard bands from topology of the ghost states}
\label{sec:TopoHubbardBands}
We now turn to the investigation of the topology associated with high-energy excitations, such as the lower (LHB) and upper (UHB) Hubbard bands of an interaction-driven MI, as captured by the quasiparticle Hamiltonian $H_\mathrm{qp}(\bk)$. To isolate this effect, we consider the simplest realization of a MI phase in the BHZ model with the interaction in Eq.\,\eqn{Hint}, setting $U_s=U_p=8$, $V=0$ and $J=0$. Under these conditions, the system reduces to two copies of the single-band Hubbard model in the Mott insulating regime, weakly coupled by the noninteracting hybridization hopping terms between the orbitals. 

In Fig.\,\ref{fig:TopologyHubbardBands} we show the band structure of $H_\mathrm{qp}(\bk)$ for three representative values of the on-site bare mass $M$, both in PBC and OBC. The resulting quasiparticle band-structure exhibits two hallmark features of the gGut description of a MI: (i) a flat band at the Fermi energy for each orbital, corresponding to the auxiliary fermionic degree of freedom that mimics the narrow metallic peak in the interacting DOS, but with vanishing dispersion, hence no weight in the physical DOS of the insulator; (ii) a pair of Hubbard bands per orbital, which encode the high-energy charge excitations of the system. Notably, for $M \lesssim 1$, the Hubbard bands associated with the two orbitals hybridize, develop a mixed orbital character, and open a topological gap. The color scale indicates the orbital polarization of each band, highlighting regions where band inversion occurs. For the latter regime also the spin-resolved Chern numbers of the ghosts-bands signal a nontrivial topology. According to the bulk-boundary correspondence, this should result in helical edge states confined within the gaps of the lower and upper Hubbard bands when the system is placed in an open geometry. In the lower panels of Fig.\,\ref{fig:TopologyHubbardBands} we confirm this by showing the eigenvalue spectrum of $H_\mathrm{qp}(k_x)$ for a slab geometry with OBC along $\hat{y}$ ($N_y = 20$) and PBC along $\hat{x}$. As expected, for $M \lesssim 1$, two pairs of edge states emerge in the high-energy topological gaps. The color scale again quantifies the spatial localization of these states at the boundaries, confirming their edge character. We note that the gap closing/opening visible in the high-energy Hubbard-like bands corresponds to a finite-energy band touching among poles of $H_{\mathrm{qp}}(\mathbf{k})$. This feature reflects the reorganization of spectral poles/weights and should not be conflated with the ground-state topological transition, which is determined by the quasiparticle gap at the Fermi level. An intriguing aspect of our results is the value for which we observe the topological transition between the Hubbard bands. Specifically, we observe that the transition occurs around $M \approx M_\mathrm{c} /2 = 1 $, where $M_\mathrm{c}$ is the critical mass that marks the topological phase-transition in the noninteracting model. We can rationalize this result by noticing that, well inside the MI, the self-energy is well approximated by 
\be
\Sigma(i \epsilon) \simeq \frac{\Delta^2}{i \epsilon - M \sigma_0 \otimes \tau_3} + \frac{U}{2} \sigma_0 \otimes \tau_0,
\ee
the first term originating from an expansion around the atomic limit with $\Delta$ proportional to the Mott gap, and the second accounting for the Hartree-Fock shift. At half-filling, this Hartree-Fock term exactly compensates the chemical potential, and the retarded single-particle Green's function reads
\be
G(\omega + i \eta, \mathbf{k} ) = \frac{1}{\omega - H_0(\mathbf{k}) + \mu - \Sigma(\omega + i \eta) + i \eta}.
\ee
Solving for the poles yields the condition
\be
\omega - H_0(\mathbf{k}) - \frac{\Delta^2}{ \omega - M \sigma_0 \otimes \tau_3} = 0,
\ee
which, at leading order in $\Delta \gg 1$, gives the Hubbard bands dispersions
\be
\omega_{\pm} = \frac{H_0(\mathbf{k}) + M \sigma_0 \otimes \tau_3}{2} \pm \Delta.
\ee
In the resulting dispersion the hopping amplitudes are reduced by a factor of two with respect to the bare dispersion ones, while $M$ remains approximately the same. This scaling explains the emergence of a topological transition in the Hubbard bands at $M \approx M_\mathrm{c} / 2$.

As highlighted in the approximate analytic argument, this shift is a genuinely nonperturbative feature of the high-energy spectrum in the Mott insulating regime, and is fully captured within the gGut framework. It is worth noting that a topological classification of MIs has also been proposed based on the structure of the zeros of the Green’s function: when nonlocal dynamical correlations are included—beyond what is accessible in single-site DMFT or gGut—dispersive zeros emerge within the Mott gap. These zeros can acquire a topologically nontrivial structure, contributing to a finite winding number of the Green’s function itself \cite{MottZerosGiorgio,AndreaTopoZeros, AndreaTopoZeros, AndreaTKI, PasquaSciPost, PangburnArXiv}. Importantly, the existence of zeros within the Mott gap reflects the underlying structure of the Green’s function as a sum of contributions from the LHB and UHB poles; the presence of a pair of poles generically implies a zero in between. This observation suggests that, at least at leading order, the topology of the Hubbard bands and the topology of Green's function zeros may originate from a common nonperturbative mechanism. Further exploration of this potential link could shed light on how local and nonlocal correlations cooperate in defining the topological structure of interacting systems. A detailed analysis will require cluster extensions of gGut that incorporate nonlocal correlation effects, which we leave for future studies.

Given the interpretability of our framework, it presents particular promise to aid in the design of novel materials with unexplored topological properties. In recent years, leveraging the interplay between topology and magnetism has emerged as a fruitful direction~\cite{Otrokov2019,Zhang2019,Bernevig2022}, and we expect the gGut formalism to provide meaningful insights in this field. 
To illustrate this, we consider the MI with topological Hubbard bands discussed in the previous section, and build upon the magnetic response of the paramagnetic Mott insulator, previously explored in~\cite{guerci2019, Tagliente2025}, to investigate how a magnetic field affects the topology of the high-energy Hubbard bands.

\begin{figure}[t!]
    \centering
    \includegraphics[width=0.45\textwidth]{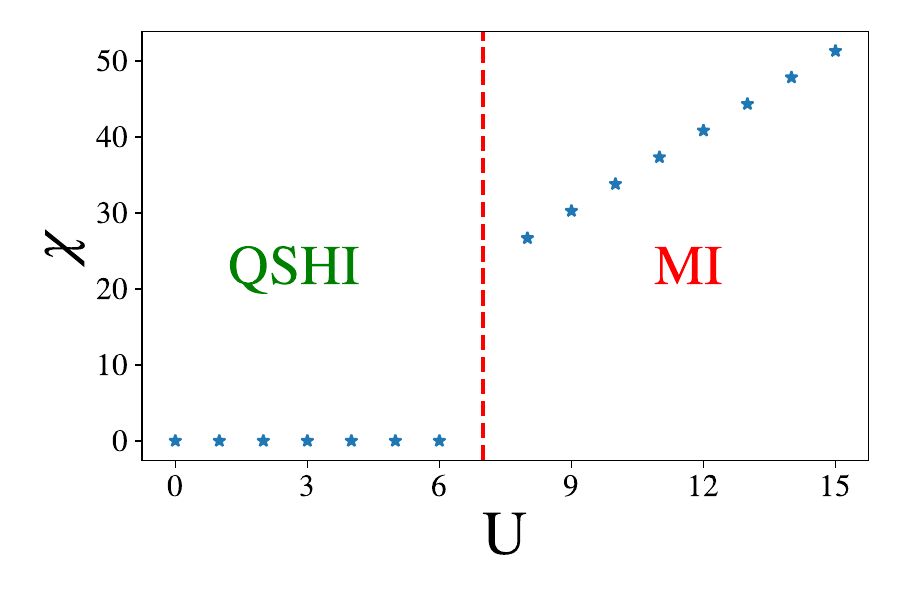}
    \caption{Magnetic susceptibility $\chi$ as a function of the interaction strength $U$, for fixed $M = 1.0$. The red dashed line marks the transition between the two insulating phases. In the QSHI phase, $\chi$ vanishes, consistent with the gapped nature of the insulating state. In contrast, the MI phase exhibits a finite susceptibility that increases as $\chi \sim U$. This paramagnetic behavior suggests the presence of gapless quasiparticles inside the gap with their own Fermi surface, see text for details.} 
    \label{fig:MIParaChi}
\end{figure}

To this end, a standard procedure is to apply a small, uniform and static magnetic field to the lattice to induce a finite magnetization at each site. The field $h$ is chosen sufficiently small to remain in the linear response regime. Thus, the magnetic susceptibility would be obtained as $\chi = \partial_h m |_{h \to 0}$. However, this procedure can fail in a MI due to the magnetic instability of the ground state \cite{guerci2019}. As the variational nature of the gGut method allows us to study solutions with fixed symmetries, we can use this to our advantage in this case. In our calculations, we access the ferromagnetic state by fixing the magnetization on the impurity site, effectively selecting a suitable linear combination of the two lowest-energy eigenstates in the impurity problem~\footnote{In the $h = 0$ limit, these are degenerate singlet and triplet configurations}. This approach is physically equivalent to studying the system in the presence of a magnetic field, while avoiding the subtlety associated with a degenerate ground state in the impurity problem. It further allows us to compute the ground state energy as a function of the magnetization $E(m)$ . It is then a matter of simple algebra to show that the magnetic susceptibility can be obtained from the curvature of $E(m)$ at $m=0$, as 
\begin{equation}
    \chi = \partial_h m|_{h\to0} = ( \partial^2_m E(m))^{-1}_{m \to 0}.
\end{equation}
Examining the magnetic susceptibility, shown in Fig.\,\ref{fig:MIParaChi}, we find that in the QSHI phase $\chi$ vanishes, consistent with the gapped nature of the insulating state. In contrast, the MI phase exhibits a finite susceptibility that increases as $\chi \sim U$. This paramagnetic behavior suggests the presence of gapless quasiparticles with an emergent Fermi surface~\cite{Tagliente2025}. Let us analyze the structure of the gGut solution in this regime.
When the MI phase acquires finite magnetization, the spin-imbalance in the impurity orbital is exactly compensated by the decoupled bath state of the impurity model, which in turn is associated with the flat band at the Fermi level in the quasiparticle Hamiltonian. Meanwhile, the other two bath sites, which correspond to the high-energy Hubbard bands, remain unpolarized. Nevertheless, their hybridization with the magnetized impurity is necessarily spin-dependent, leading to $R$-matrix elements for the Hubbard bands which are different for spin-up and spin-down electrons. In other words, the dispersion of the quasiparticle Hubbard bands develops a spin-dependent renormalization. This occurs in such a way that the spin-up dispersion of the upper Hubbard band exactly matches the spin-down dispersion of the lower Hubbard band, and vice versa. 
\begin{figure}[t!]

    \centering
    \includegraphics[width=0.50\textwidth]{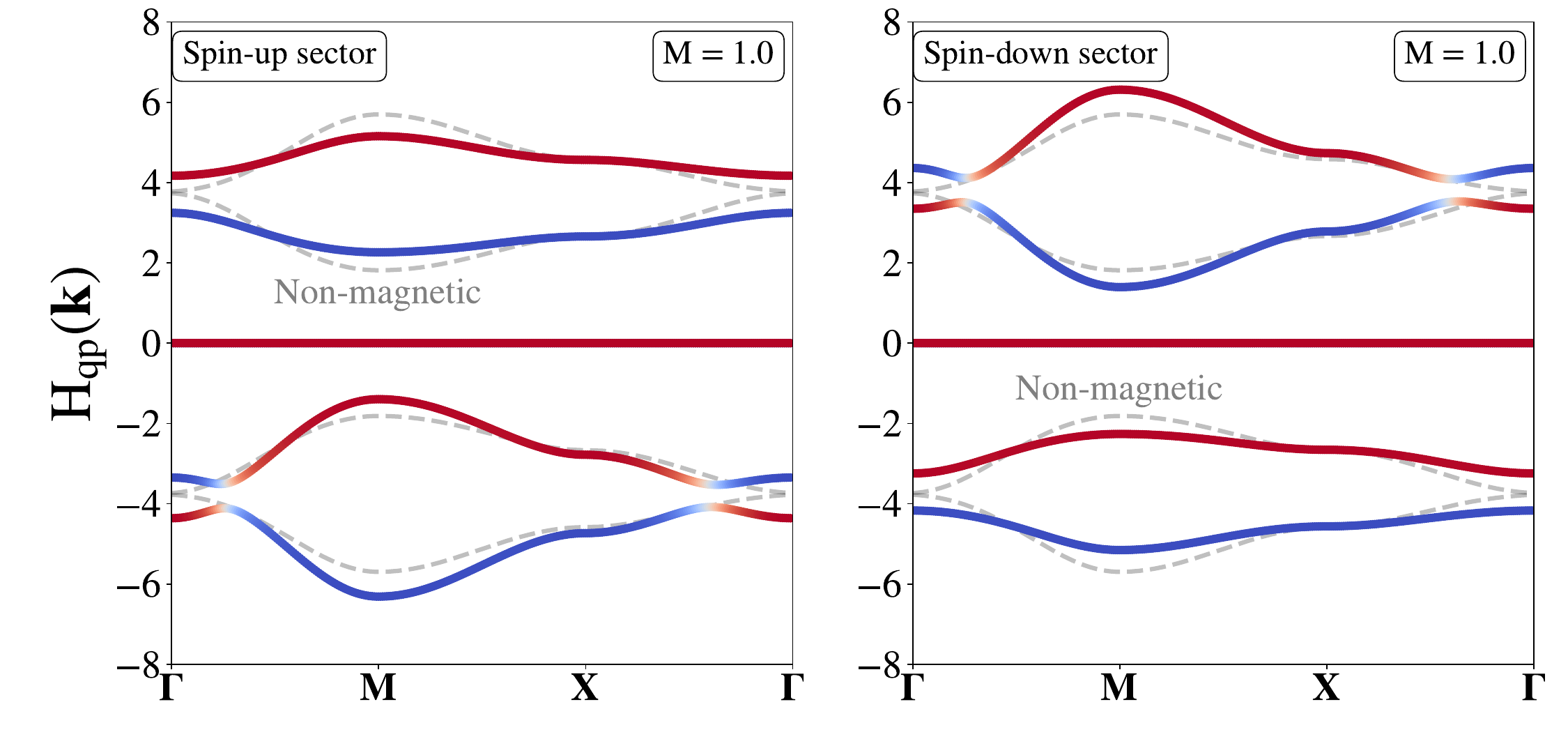}
    \caption{Quasiparticle band structure of $H_{\mathrm{qp}}(\bk)$ showing the renormalization of the Hubbard bands in presence of a finite magnetization $m = 0.2$ per orbital, for $U = 8.0$ and $M=1.0$. The color scale indicates the orbital polarization (red for $s$ and blue for $p$ orbital). Gray dashed lines indicate the corresponding band structure in the absence of magnetization. In the spin-up sector, the lower Hubbard bands acquire a non-zero Chern number, while the upper Hubbard bands remain topologically trivial. The opposite occurs in the spin-down sector. }
    \label{fig:FMCherniMI}
\end{figure}
\section{Interplay between topology, magnetism and correlation}

To understand the origin of this phenomenon, the physical electrons can be considered as composed of multiple quasiparticles. Indeed, the renormalization parameter $R$ can be understood as describing how each physical electron is fractionalized into several quasiparticles within the gGut description, as one way of arriving at $H_\mathrm{qp}$ in Eq.\,\eqref{eq:Hqp} is substituting $c^\dagger_{\alpha}\sim\sum_a R^\dagger_{a\alpha} d^\dagger_a$. In this sense, the Hubbard quasiparticle bands share the charge density of the physical band, even if its spin is carried in its entirety by the flat quasiparticle band at the Fermi level. When the physical band is exactly half-filled, both upper and lower Hubbard bands have an equal distribution of the physical electron charge. Yet, when the physical band becomes magnetized, say in positive $S_z$ direction, the spin-up physical band becomes more than half-filled. Consequently, the spin-up lower Hubbard quasiparticle band becomes its larger contributor, since it is below the Fermi level and thus filled, and sees its $R$ coefficient increased. Meanwhile, the $R$ coefficient of the spin-up upper Hubbard band is decreased. The opposite effect is observed for the spin-down quasiparticle bands, as the spin-down physical band becomes less than half-filled. In simpler terms, the physical magnetization is not enforced by modifying the occupation of the \textit{quasiparticle} Hubbard bands, but rather by changing their weights in the physical electron excitation. Furthermore, the above discussion highlights the role---here driven by the magnetization---of this decoupled flat band in determining the dispersion of the Hubbard bands. A similar mechanism has been observed in \cite{Stepanov2024}.

In essence, we see that the gGut quasiparticle description of the magnetized MI can be understood in terms of an effective spin-charge separation of the physical electrons, in which the spin component is carried by the decoupled flat quasiparticle band, while the charge component is provided by the quasiparticle Hubbard bands. 

Returning to the interplay of topology and magnetism, we specifically leverage the renormalization of the Hubbard band dispersion---induced by spin-dependent $R$-matrix elements in the presence of finite magnetization---to propose a new mechanism by which the high-energy topological Hubbard bands evolve into spin-selective Chern bands. In this scenario, only one spin species exhibits a non-zero and quantized Chern number. As shown in Fig.\,\ref{fig:FMCherniMI}, for a magnetization $m=0.2$ per orbital, the QP bands in the spin-up sector are renormalized such that the lower Hubbard bands have a band crossing and subsequent gap opening, which can host chiral edge states in OBC, though at finite energy. This follows from the increase of the $R$ coefficient in this band which accounts for the excess spin-up charge in the physical sites. In contrast, the corresponding upper Hubbard bands in the same spin sector renormalize in the opposite direction and remain topologically trivial. The QP bands in the spin-down sector exhibit the complementary behavior: the upper Hubbard bands develop a non-trivial topology while the lower bands remain trivial. As a result, the topological structure of the Hubbard bands in this MI phase becomes spin-selective, with a quantized Chern number emerging only for one of the spins.
We see thus how the quasiparticle picture in gGut enables simple, single-particle based descriptions of correlated topology and magnetism, which one can leverage to engineer exotic phases of matter, as well as for examining predictions done at less correlated levels of theory.

\section{Conclusion}
We have shown how the quasiparticle description at the core of the ghost Gutzwiller (gGut) variational ansatz allows bridging two foundational pillars of modern condensed matter physics: band structure topology, rooted in a single-particle description, and strong electronic correlations, which typically undermine such a picture. Working on the paradigmatic interacting BHZ model, we have shown that the quasiparticle Hamiltonian in gGut codifies the topology of the original correlated system, and that it can be used not only for a computationally efficient phase space exploration, but also to model the interplay of topology and correlation with a heretofore unreached level of interpretability, especially for states sitting far from the Fermi level. 

A clear demonstration of this interpretability arises in the analysis of the Chern numbers associated with the quasiparticle bands. These quantities offer a transparent and quantitatively accurate interpretation of the Green’s function winding number, currently the state-of-the-art method for characterizing the integrated topology in correlated systems. A major innovation of the gGut framework, however, lies in its ability to capture the topology of high-energy excitations, such as the Hubbard bands, thereby extending the analysis to energy- and momentum-resolved topological features. This enables not only a more precise comparison with experimentally accessible ARPES spectra, as already pointed out in \cite{Lee2023a}, but also the identification of genuinely nonperturbative features that lie beyond the reach of conventional DFT- and topological Hamiltonian-based methods. More broadly, we have shown that the gGut formalism promotes a physically grounded interpretation of the Green’s function through its quasiparticle structure, offering a step forward in the theoretical understanding of correlated topological phases and their interplay with magnetism, and complementing contemporary strategies to describe strong correlation and band structure topology comprehensively~\cite{Barry2025}.

Overall, our study paints a convincing portrait of the gGut framework as a reliable, efficient and, most importantly, 
interpretable tool to study the interplay between correlation and topology.
By restoring a quasiparticle perspective within a correlated setting, it brings the description of many-body phenomena 
closer to the language of band structure theory, unlocking the potential for predictive modeling in correlated materials. 
This capability paves the way to extend powerful paradigms, such as topological quantum chemistry 
\cite{Slager2013,Kruthoff2017,Bradlyn2017,Antimo_topoDFT,Vergniory2019}, to strongly interacting regimes
\cite{Lanata2017,Lee2023a,Lee2023b,Mejuto2023a,Mejuto2024,Giuli2025,Tagliente2025,Bellomia_intracorr}, beyond the 
traditional domain of weakly correlated systems. Moreover, it opens the door to high-throughput computational searches 
of correlated topological materials, accelerating the discovery of novel platforms for low-power electronics, spintronics, 
and quantum memory devices.

\section*{Acknowledgments}
We acknowledge insightful discussions with A.~Amaricci, M.~Capone, N.~Lanat\`a, A.~Marrazzo, R.~Resta and J.~Skolimowski. I.P. also gratefully acknowledges S.~Giuli for valuable assistance with the numerical implementation during the initial stages of the project.
G.B. is financially supported by the National Recovery and Resilience Plan PNRR
MUR Project No.~CN00000013-ICSC and by MUR via the PRIN 2020 (Prot.~2020JLZ52N-002) and 
PRIN 2022 (Prot.~20228YCYY7) programs. 
B.M. acknowledges support from a UKRI Future Leaders Fellowship [MR/V023926/1] and from the Gianna Angelopoulos Programme for Science, Technology, and Innovation. 

\appendix
\section{ghost Gutzwiller Equations}

In this section, we briefly summarize the main equations of the ghost Gutzwiller (gGut) method within its embedding formulation.
For derivations or discussions on the implementation, we refer to the existing literature and references therein~\cite{Fabrizio2007,lanata2015,Lanata2017,Mejuto2023a}.
We will focus here on the application of gGut to lattice models with exclusively local interactions, implementations for finite-size systems with nonlocal interactions can be found in Ref.~\cite{Mejuto2024}.
The physical Hamiltonian can be written in this case as a noninteracting, lattice contribution and a sum of local interaction terms, as
\beal
    H_\mathrm{phys} &= \sum_I H_I^\mathrm{loc} + H_\mathrm{latt},\\
    H_I^\mathrm{loc} &= \sum_{\alpha\beta} t^{II}_{\alpha\beta}\ c^\dagger_{I\alpha}c^\dagga_{I\beta}+\frac{1}{2}\sum_{\alpha\beta\gamma\delta} U_{\alpha\beta;\gamma\delta}\ c^\dagger_{I\alpha}c^\dagger_{I\gamma}c^\dagga_{I\delta}c^\dagga_{I\beta},\\
    H_\mathrm{latt} &= \sum_{I\neq J}\sum_{\alpha\beta} t^{IJ}_{\alpha\beta}\ c^\dagger_{I\alpha}c^\dagga_{J\beta} = \sum_{\bk}\sum_{\alpha\beta}\epsilon^{\alpha\beta}_\bk c^\dagger_{\bk\alpha}c^\dagga_{\bk\beta}.
    \label{eq:SI_Hphys}
\eal
Here, $c^\dagger$/$c$ indicate creation/annihilation operators in the physical space, Greek indices run over the physical orbitals (potentially including spin), $t^{IJ}_{\alpha\beta}$ corresponds to the one-body Hamiltonian term between atoms $I$ and $J$ and orbitals $\alpha$ and $\beta$, and the tensor $U$ represents the local interactions.
In the last line, we have used the space translational invariance to write the one-body lattice Hamiltonian in terms of the momentum dependent dispersion $\epsilon^{\alpha\beta}_\bk$.
Within gGut, the ground state of $H_\mathrm{phys}$ is obtained in terms of a self-consistent embedding between two model representations of the system: a local, interacting impurity Hamiltonian $H_\mathrm{imp}$ and an extended, noninteracting quasiparticle Hamiltonian $H_\mathrm{qp}$.
The latter follows the expression in Eq.\,\eqref{eq:Hqp} in the main text, namely
\beal
    H_\mathrm{qp} = \sum_{\bk}\sum_{ab}\left[\sum_{\alpha\beta}R^\dagger_{a\alpha}\epsilon^{\alpha\beta}_\bk R^\dagga_{\beta b}-\lambda_{ab}\right]d^\dagger_{\bk a}d^\dagga_{\bk b},
    \label{eq:SI_Hqp}
\eal
where $d^\dagger$/$d$ identify quasiparticle creation/annihilation operators, Latin letters run over the quasiparticle orbitals (potentially including spin), and we have introduced the main gGut parameters to be determined self-consistently: $R$ and $\lambda$.
The $R$ matrices represent the renormalization of the noninteracting dispersion, while $\lambda$ is a local, one-body potential.
These are introduced to represent the effects of electronic correlation within an effective, band structure theory.
Note that the number of quasiparticle orbitals can be larger than the number of physical orbitals once ghosts are included, making the matrices $R$ rectangular.
It is this inclusion which allows gGut to capture high-energy features of correlation within an effective quasiparticle description~\cite{Lanata2017,Mejuto2023a}.

On the other side of the self-consistency we have the local impurity Hamiltonian $H_\mathrm{imp}$, which follows
\beal
    H_\mathrm{imp} = H^\mathrm{loc}_0+\sum_{\alpha a}\left(V_{a\alpha}d^\dagger_{a}c^\dagga_\alpha+\mathrm{h.c.}\right)-\sum_{ab}\lambda^\mathrm{c}_{ab}d^\dagga_{b}d^\dagger_a,
    \label{eq:SI_Himp}
\eal
where $H^\mathrm{loc}_0$ corresponds to the local part of the physical Hamiltonian for one of the atoms, expressed in terms of $c^\dagger_{\alpha}\equiv c^\dagger_{0\alpha}$.
This single atom is hybridized with a finite bath of quasiparticle orbitals $d^\dagger$, through the couplings $V_{a\alpha}$ and the local bath potential $\lambda^\mathrm{c}_{ab}$.
As will be shown below, these bath parameters follow directly from the solution of the quasiparticle Hamiltonian.
Both Hamiltonian descriptions are connected through the self-consistency condition in gGut, which imposes that the one-body reduced density matrices of $H_\mathrm{imp}$ and $H_\mathrm{qp}$ be related by
\beal
    \langle d^\dagga_b d^\dagger_a\rangle_\mathrm{imp}  \overset{!}{=} \langle d^\dagger_{0a} d^\dagga_{0b}\rangle_\mathrm{qp} \equiv \Delta^\mathrm{qp}_{ab}.
    \label{eq:SI_scf}
\eal

Essentially, $H_\mathrm{imp}$ captures the local correlation of the lattice model, which is fed through the self-consistency into the quasiparticle model in an enlarged Hilbert space. 
This self-consistent embedding method is equivalent to a variational energy optimization within the infinite dimensional limit.

Practically, the self-consistency is solved iteratively, starting from some guess $R$ and $\lambda$ and going over the following steps

\begin{enumerate}
    \item The quasiparticle Hamiltonian $H_\mathrm{qp}$ is formed using the current $R$ and $\lambda$ according to Eq.\,\eqref{eq:SI_Hqp}. Its ground state local one-body reduced density matrix $\Delta^\mathrm{qp}_{ab} = \langle d^\dagger_{0a} d^\dagga_{0b}\rangle_\mathrm{qp}$ is evaluated.
    \item The impurity model bath parameters $V$ and $\lambda^\mathrm{c}$ can then be evaluated following
    \beal
        V &= \frac{1}{V_\mathrm{BZ}}\left[\sqrt{\Delta^\mathrm{qp}\cdot\left(\mathbb{I}-\Delta^\mathrm{qp}\right)}\right]^{-1}\cdot\sum_{\bk} \Delta^\mathrm{T}_{\bk}\cdot R^\dagger \cdot \epsilon_\bk,\\
        \lambda^\mathrm{c}_{ab} &= -\lambda_{ab} + \frac{\partial}{\partial\Delta^\mathrm{qp}_{ab}}\mathrm{Tr}\left[R\cdot\sqrt{\Delta^\mathrm{qp}\cdot\left(\mathbb{I}-\Delta^\mathrm{qp}\right)}\cdot V\right]+\mathrm{h.c.},
        \label{eq:SI_VandLc}
    \eal
    where $V_\mathrm{BZ}$ is the volume of the Brillouin zone of the quasiparticle model, $\Delta_{\bk;ab}=\langle d^\dagger_{\bk a}d^\dagga_{\bk b}\rangle$ and the $\cdot$ denotes matrix products in orbital space.
    The partial derivative in the equation for $\lambda^\mathrm{c}$ acts only on the square-root term inside the trace.
    \item The computed $V$ and $\lambda^\mathrm{c}$ determine the impurity model in Eq.\,\eqref{eq:SI_Himp}, whose ground state one-body density matrix is evaluated. In particular, we are interested in the $\langle c^\dagger_\alpha d^\dagga_a \rangle_\mathrm{imp}$ and $\langle d^\dagger_a d^\dagga_b\rangle_\mathrm{imp}$ components. In this work, we employ an exact diagonalization solver.
    \item Finally, we can close the iteration by proposing a new pair $R$, $\lambda$ from the impurity model density matrix, according to
    \beal
        \mathbb{I}-\langle d^\dagger_a d^\dagga_b\rangle_\mathrm{imp} &= \Delta^\mathrm{qp}_{ab},\\
        \langle c^\dagger_{\alpha} d^\dagga_b\rangle_\mathrm{imp} &= \sum_a R_{\alpha a}\left[\sqrt{\Delta^\mathrm{qp}\cdot\left(\mathbb{I}-\Delta^\mathrm{qp}\right)}^\mathrm{T}\right]_{ab}.
    \eal
    \item The iteration is repeated from step 1 until the $R$ and $\lambda$ matrices are converged within some tolerance. To speed up convergence, we employ a DIIS algorithm~\cite{Shepard2007} to propose new parameters in each iteration.
\end{enumerate}

Given the spatial translational symmetry in our periodic lattices, we typically only solve a single impurity model per iteration.
In the multi-layered configurations, we represented each layer by an independent impurity problem, all of which couple through the quasiparticle Hamiltonian.


%

\end{document}